\documentclass[final,1p,times,twocolumn]{elsarticle}
\usepackage{graphicx}
\usepackage{amssymb}
\usepackage{amsthm}

\usepackage{tabularx}
\usepackage{multirow}

\journal{ASHRAE Journal}

\begin{document}

\begin{frontmatter}



\title{Estimating electricity saving-potential in small offices using adaptive thermal comfort}

\author[mainaddress]{Yujiao Chen}
\author[address2]{Rongxin Yin\corref{correspondingauthor1}}
\cortext[correspondingauthor1]{Corresponding author: Rongxin Yin}
% \cortext[correspondingauthor1]{Corresponding author: Rongxin Yin, ryin@lbl.gov}

\ead{ryin@lbl.gov}
\address[mainaddress]{Sidewalk Labs, New York, NY, USA}
\address[address2]{Lawrence Berkeley National Laboratory, Berkeley, CA, USA}
% \author[Rongxin Yin]{Rongxin Yin\corref{correspondingauthor1}}

\begin{abstract}

Small office buildings below 10,000 square feet account for approximately 75\% of the total number of office buildings in the United States. Unlike large office buildings, small office buildings usually use packaged HVAC systems for environmental control and are not equipped with building management systems. The lack of smart control in small office buildings leaves low-hanging fruit unpicked in terms of energy efficiency. The commonly seen missed opportunities include the fixed cooling setpoint schedule that ignores people’s comfort needs and actual occupancy patterns, and the wasted energy from equipment and appliances while not being used by occupants.

The smart control informed by IoT sensors and enabled by remotely controlled devices can optimize the building operation to minimize unnecessary energy consumption and improve indoor thermal comfort. This paper quantifies the potential for electricity savings in small office buildings from smart thermostat control and occupancy-informed smart plug control. This is done by simulating the effect of adaptive setpoint temperature, occupancy-based HVAC control, and night-purge free cooling on small office buildings across all major climate zones in the United States. Adopting these smart control measures can achieve 8.9\% to 20.4\% of savings in total electricity consumption of small office buildings, or equivalent to annual reductions between 12.2 $kWh/m^2$ and 30.4 $kWh/m^2$ in electricity usage intensity. Among all climate zones, the hot and dry climates benefit the most from proposed smart controls and achieve the highest percentages of electricity savings. 
\end{abstract}

\begin{keyword}
energy efficiency \sep adaptive thermal comfort \sep IoT sensing and control \sep occupancy-based HVAC and plug load control \sep smart thermostat control

\end{keyword}

\end{frontmatter}

% \linenumbers

\section{Introduction}
\label{S:1}
In the United States, more than 73\% of electricity is consumed by buildings \cite{EnergyInformationAdministration2021Electric2019}. Electricity used for space cooling in buildings reached 392 billion kWh in 2020, which accounted for approximately 10\% of overall electricity consumption \cite{Annual2021}. Among all building types, electricity used for space cooling in commercial buildings accounted for 4\% of overall electricity consumption in the U.S. Small office buildings, with a floor area of less than 10,000 square feet, account for approximately 75\% of the total number of office buildings \cite{Annual2021}. Reducing the energy required for space cooling not only makes a great impact on total energy consumption but also significantly alleviates the stress on electricity generation and the power grid, which consequently leads to more economical operation and lower carbon intensity. 

For small office buildings, the most widely used space cooling system is packaged air-conditioning equipment, which serves more than half of the total square footage of commercial buildings in the United States \cite{EnergyData}. Packaged air-conditioning equipment usually has discrete on-off control for single-stage or two-stage cooling, operating on a fixed schedule of desired indoor air temperature (setpoints) by day and time. A more sophisticated approach to control, using commercial off-the-shelf devices and cloud-based optimization techniques can substantially improve energy efficiency and occupant satisfaction.

This paper focuses on three opportunities for improving energy use in small office buildings:

\subsection{Overcooling in office buildings}
It is not uncommon for people in the United States to experience excessively cooled office buildings on a hot summer day. Often, people need to put on an additional layer of clothes if sitting at their desk for extended periods. This is especially common for women, as described in \cite{Karjalainen2007GenderEnvironments} and \cite{Kingma2015EnergyDemand}. These studies find that females prefer higher room temperatures than males, based on large scale field surveys and actual metabolic rate through biophysical analysis. The empirical rules to set cooling temperature follow a model developed in the 1960s that derived from male test subjects only \cite{2010ASHRAEOccupancy}, which causes the intrinsic mismatch in providing comfort to all occupants. In addition to discomfort, according to \cite{Derrible2015TheStates}, the overcooling in the United States resulted in an annual energy waste of 103,929 GWh, a carbon footprint of 57,125 kt $CO_2$e, and a financial loss of 10 billion dollars. 

The technical barrier to the flexible cooling setpoint that reflects the true occupants’ preference usually lays on the lack of feedback channel and access to the setpoint control. Namely, occupants don’t have a convenient way to effectively express their thermal feedback, and there is no mechanism to collect occupants’ feedback and take appropriate actions to address it. This issue has started to improve with the implementation of consumer-grade IoT sensors and integrated smart control. Some smart building solutions, such as Mesa \cite{MesaLabs}, provide comfort buttons and a user app in their offering that allows each occupant’s thermal preference to be heard by the smart control engine, and subsequently adjusts the setpoint to improve overall occupants’ satisfaction towards their thermal environment.

\subsection{Occupancy-decoupled operation}
Programmable thermostats are the most prevailing devices to control HVAC operation in small and medium-sized offices. Those devices adjust the temperature according to a series of settings to be triggered by the time of the day and day of the week. For example, the thermostat can be programmed to set the temperature at 22°C during business hours and switch to a setback of 28°C at night and on weekends. However, the programmable thermostat itself does not save energy; its energy-efficiency performance can only be as good as how users operate it. Issues come up in practice as the previous survey results revealed that 40\% of users do not operate their programmable thermostats as expected by the manufacturers \cite{Pritoni2015EnergyBehavior}. Assumptions about business hours may fail to reflect the true occupancy -  for example, the AC turns off at 6 pm sharp while people are still working in the office, or the AC operates on Saturday when the office is unoccupied. Whoever sets the programs of thermostat in small and medium-sized office also needs a better solution - more than often, it is either the case that the thermostat is physically locked into a plastic box to prevent unauthorized interference, or multiple people interacting with the thermostat and their conflicts inadvertently cause the increase of energy use \cite{Sintov2019ThermostatBehavior}. 

The motion-sensor embedded or coupled thermostat, on the other hand, captures the true occupancy patterns instead of solely relying on user settings. This mechanism allows temperature control to be more accurate and dynamic, thus improving the energy efficiency of HVAC operations.

The occupancy-centric operation also saves energy on plug loads and lighting. It has been supported by multiple studies that plug loads account for about 30\% of electricity consumption in U.S. commercial buildings \cite{SheppyOfficeSheet,Huang2021AnStates,Hafer2015InventoryingCampus}. Unlike lighting that has widely adopted occupancy sensors for automatic control, equipment and appliances in general solely rely on manual control and consume energy even when not being used. This standby power consumption is called phantom load or vampire load. The smart plugs and power strips that allow remote control consequently save a substantial amount of energy when coupled with occupancy signals.

\subsection{Unused free cooling resources}
Most climate zones in the United States offer the opportunity to utilize natural ventilation as free cooling for various lengths of time \cite{Chen2017InvestigatingVariations}. The fresh outdoor air at the pleasant temperature and humidity level not only improves indoor air quality by flushing out the pollutant in the stale air, but also saves significant amounts of cooling and fan energy \cite{Chen2018OptimalLearning}. The most convenient implementation of natural ventilation is the night purge, which is to leave the windows and louvers open during the night if the weather is suitable. This passive ventilation is most effective in the climate with substantial diurnal temperature swings and in buildings with a significant amount of thermal mass (e.g. concrete, brick, phase-change material) \cite{Solgi2016CoolingVentilation}. 

With the increasing adoption of window actuators, IoT sensors, and integrated intelligent control, more natural ventilation opportunities can be unlocked to reduce the load and carbon footprint of building operations.

To address these missed opportunities and harvest the low-hanging fruit, this paper is intended to quantify the cooling energy savings potential in small and medium-sized office buildings across all major climate zones in the United States. By adopting adaptive cooling setpoints, occupant-centric thermostat control, and night purge as free cooling, the total cooling energy savings in kilowatt-hours are estimated for typical small and medium-sized office buildings.

\section{Methodology}
\label{S:2}
\subsection{Overall Framework}
\begin{figure}[htbp]
\centerline{\includegraphics[width=1.0\columnwidth]{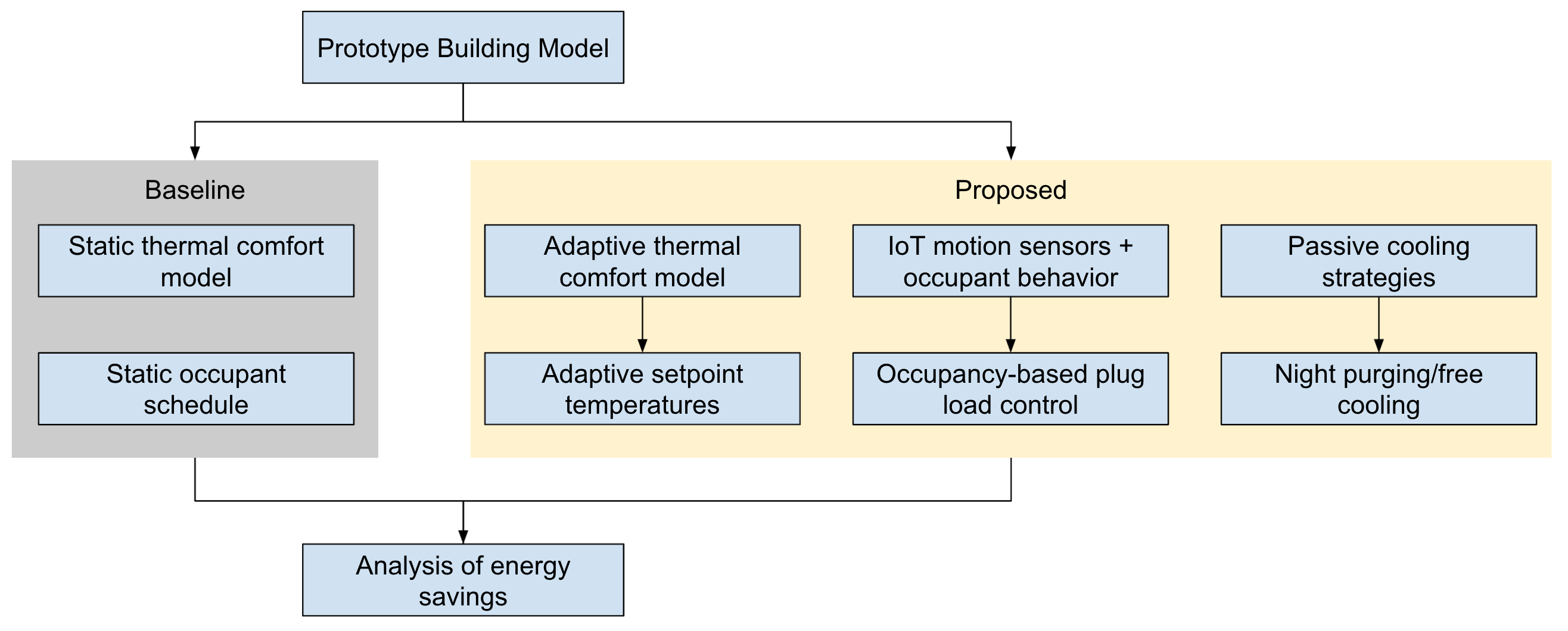}}
\caption{Proposed Modeling Framework}
\label{fig1}
\end{figure}

For each prototype building model at each climate, we will evaluate the energy-saving potentials with individual or combined measures. As illustrated in Figure \ref{fig1}, we use the DOE prototype small office model as the baseline and create the proposed model with individual and combined energy efficiency measures in EnergyPlus \cite{EnergyPlusEnergyPlus}. It is worth noting that the energy efficiency measures proposed in this study can be deployed through a set of "plug and play" IoT sensing and control technologies. Therefore, we need to keep the same HVAC system design capacity between the baseline and proposed models.

\subsection{Prototype Building Model}

The models of small office buildings to quantify cooling energy savings are based on the U.S. Department of Energy commercial benchmark models developed by DOE national labs \cite{CommercialModels}. These benchmark models serve as a consistent baseline to compare the efficiency and effectiveness of different improvements on building systems and operations. Table \ref{tab1} presents the general model input for a small office that complies with ASHRAE 90.1-2004 and 62.1-1999 standards.

\begin{table}[htbp]
\caption{Small Office Prototype Model Summary}
\begin{center}
\begin{tabularx}{1.0\textwidth} { 
  | >{\raggedright\arraybackslash}X 
  | >{\centering\arraybackslash}X 
  | >{\raggedleft\arraybackslash}X | }
\hline
\textbf{Building Type}               & \textbf{Small Office}                                                                 \\ \hline
\textbf{Compliance}                  & ASHRAE Standards 90.1-2004 and 62.1-1999                                                \\ \hline
\textbf{Description}                 & Single-story five-zone office building                                                \\ \hline
\textbf{Total Floor Area}            & 511 m$^2$ (5,500 ft$^2$)                                                                    \\ \hline
\textbf{HVAC - Cooling}              & Packaged Single Zone Air Conditioner (PSZ-AC)                                         \\ \hline
\textbf{Economizer}                  & No                                                                                    \\ \hline
\textbf{Internal Gains - Lighting}   & 1 W/ft$^2$                                                                               \\ \hline
\textbf{Internal Gains - Plug Loads} & 1 W/ft$^2$                                                                               \\ \hline
\textbf{Internal Gains - People}     & 5 W/ 1,000 ft$^2$                                                                         \\ \hline
\textbf{Envelope}                     & Envelope thermal properties vary with climate according to ASHRAE Standard 90.1-2004. \\ \hline
\end{tabularx}
\label{tab1}
\end{center}
\end{table}

\subsection{Climate zone}

The climate zone designation is based on the climate zone map used by the International Energy Conservation Code (IECC) and the American Society of Heating, Refrigerating and Air-Conditioning Engineers (ASHRAE). This climate zone map was developed by Pacific Northwest National Laboratory, dividing the United States into eight main climate zones based on the temperature, using cooling degree days and heating degree days. Ordered by descending average temperature, these zones are Zone 1 (Very Hot), Zone 2 (Hot), Zone 3 (Warm), Zone 4 (Mixed), Zone 5 (Cool), Zone 6 (Cold), and Zone 7 (Very Cold), and Zone 8 (subarctic). Each zone is further divided into three subcategories by moisture regimes, as A (Humid), B (Dry), and C (Marine) \cite{Baechler2015GuideCounty}.

The land and population under each climate zone vary dramatically in the United States. For example, Zone 8 Subarctic climate is only found in Alaska. The number of small and medium-sized office buildings under each climate zone, summarized by researchers at Pacific Northwest National Laboratory and National Renewable Energy Laboratory \cite{Baechler2015GuideCounty} is listed in Table \ref{tab2}.

\begin{table}[htbp]
\caption{Number of Buildings by Type and Climate Zone in the United States}
\begin{center}
\begin{tabularx}{1\textwidth} { 
  | >{\raggedright\arraybackslash}X 
  | >{\centering\arraybackslash}X 
  | >{\raggedleft\arraybackslash}X | }
\hline
\textbf{Climate Zone}                   & \textbf{Represent City}   & \multicolumn{1}{l|}{\textbf{\# Small Office}} \\ \hline
1A (Hot-Humid)               & Miami, Florida            & 28,000                                        \\ \hline
2A (Hot-Humid)                & Houston, Texas            & 354,930                                       \\ \hline
2B (Hot-Dry)                  & Phoenix, Arizona          & 96,470                                        \\ \hline
3A (Hot-Humid and Mixed-Humid) & Atlanta, Georgia          & 321,170                                       \\ \hline
3B (Hot-Dry)                  & Las Vegas, Nevada         & 158,390                                       \\ \hline
3C (Marine)                   & San Francisco, California & 25,930                                        \\ \hline
4A (Mixed-Humid)             & New York, New York        & 312,150                                       \\ \hline
4B (Mixed-Dry)                & Albuquerque, New Mexico   & 15,790                                        \\ \hline
4C (Marine)                   & Seattle, Washington       & 40,880                                        \\ \hline
5A (Cold)                               & Chicago, Illinois         & 306,880                                       \\ \hline
5B (Cold)                               & Boulder, Colorado         & 107,340                                       \\ \hline
6A (Cold)                               & Minneapolis, Minnesota    & 80,460                                        \\ \hline
6B (Cold)                               & Helena, Montana           & 10,080                                        \\ \hline
7 (Very Cold)                           & Duluth, Minnesota         & 10,790                                        \\ \hline
\end{tabularx}
\label{tab2}
\end{center}
\end{table}

% \subsection{Energyplus}
% EnergyPlus \cite{EnergyPlusEnergyPlus} is a physics-based building simulation program developed by DOE national labs to model thermodynamics processes and energy consumption within the building. It has the whole-building simulation capacity for space heating and cooling, mechanical and mixed-mode ventilation, and energy use on lighting and plug loads. 

\subsection{Adaptive Thermal Comfort Model}
The adaptive thermal comfort model was developed through a data-driven approach from the ASHRAE Global Thermal Comfort Database \cite{ASHRAEII}. The adaptive comfort model relates the zone operative temperature setpoint to the recent history of outdoor air temperatures. The adaptive comfort models only apply to cooling mode and generate a single setpoint value for each day. During summertime in hot climates, the zone thermostat setting can be higher than the traditional thermostat setting based on the adaptive thermal comfort model, which results in energy savings of HVAC systems.

\begin{figure}[htbp]
\centerline{\includegraphics[width=1.0\columnwidth]{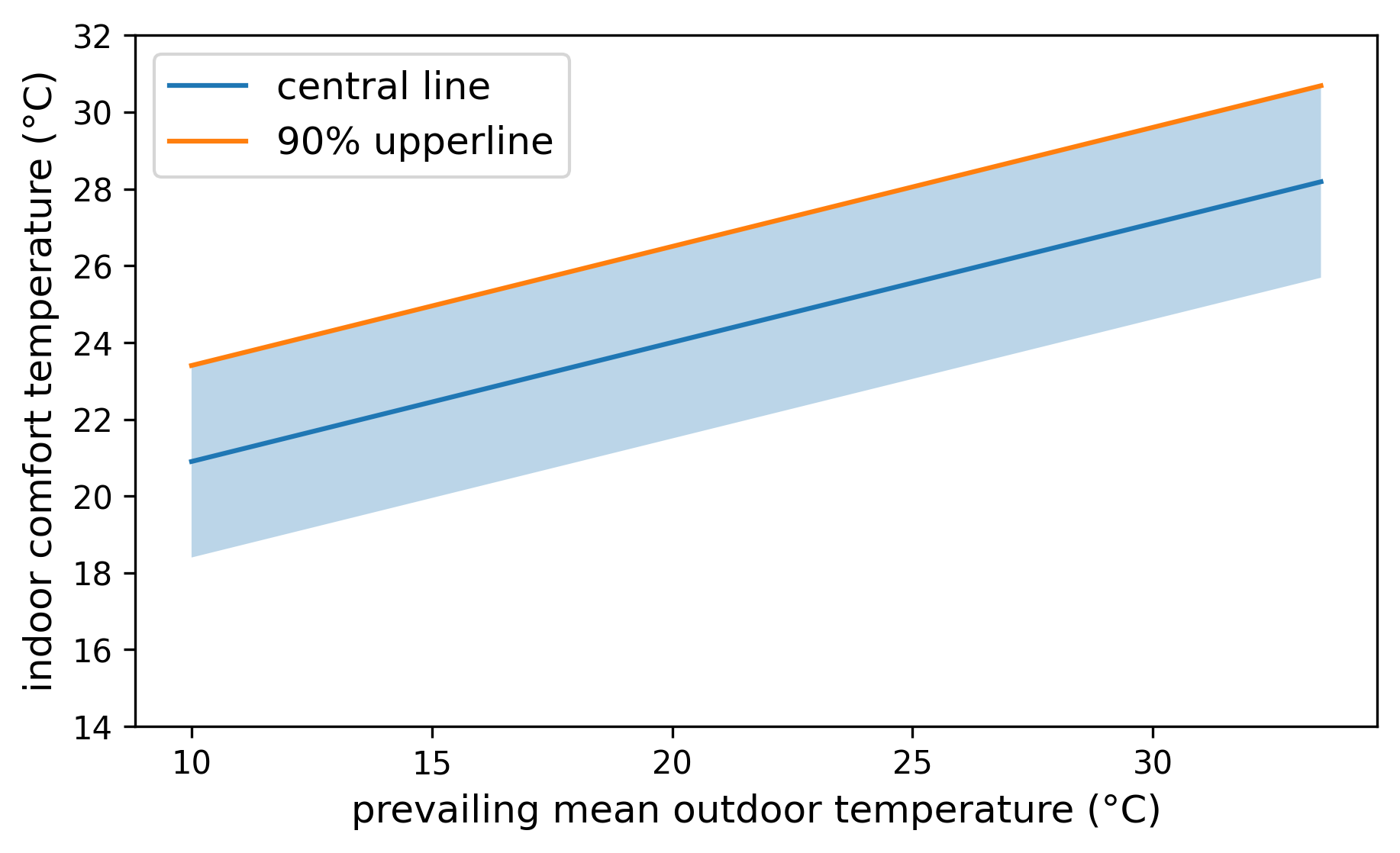}}
\caption{ASHRAE 55 Adaptive Comfort Model}
\label{fig13}
\end{figure}

For the adaptive comfort model selected in this paper, the thermostat setpoint temperature schedule for space cooling will be overwritten with the calculated operative temperature based on the selected acceptability limits of the comfort model defined in ASHRAE 55-2020 \cite{2010ASHRAEOccupancy}, as shown in Figure \ref{fig13}
. The adaptive comfort model takes into account the variation of comfort preference along with the seasonality and has a wider comfort range. The ASHRAE adaptive comfort model is only applicable when the running average outdoor air temperature for the past 30 days is between 10.0 and 33.5°C.

\section{Results}
\label{S:3}

\subsection{Adaptive Thermal Comfort Model}
In this study, we first compare the temperature range between the ASHRAE 55 standard and adaptive comfort model as follows.

\begin{enumerate}
  \item "Adaptive-ASH55-Central" (Adaptive-55) is the central line of the acceptability limits of the ASHRAE Standard 55-2020 adaptive comfort model that will be used to generate the zone operative temperature setpoint. It is derived from the prevailing mean outdoor temperature and is calculated as shown below Central line = 0.31 prevailing mean outdoor temperature + 17.8°C The Adaptive-55 is strictly bounded between 20.9°C and 28.2°C.
  
  \item "Adaptive-ASH55-90percent-Upper" (Adaptive-90) is the upper line of the 90\% acceptability limits of the ASHRAE Standard 55-2020 adaptive comfort model. It is calculated as the equation shown below
    90\% upper line = 0.31 prevailing mean outdoor temperature + 20.3°C
    The Adaptive-90 is strictly bounded between 23.4°C and 30.6°C. This upper line is only used in this section to demonstrate additional cooling energy savings compared to the Adaptive-55. In all other sections, only the Adaptive-55 is used to quantify the energy savings. 
    
  \item The adaptive comfort models only apply to cooling mode and generate a single setpoint value for each day. During summertime in hot climates, the zone thermostat setting can be higher than the traditional thermostat setting based on the adaptive thermal comfort model, which results in energy savings of HVAC systems.
\end{enumerate}

\begin{figure}[htbp]
\centerline{\includegraphics[width=1.0\columnwidth]{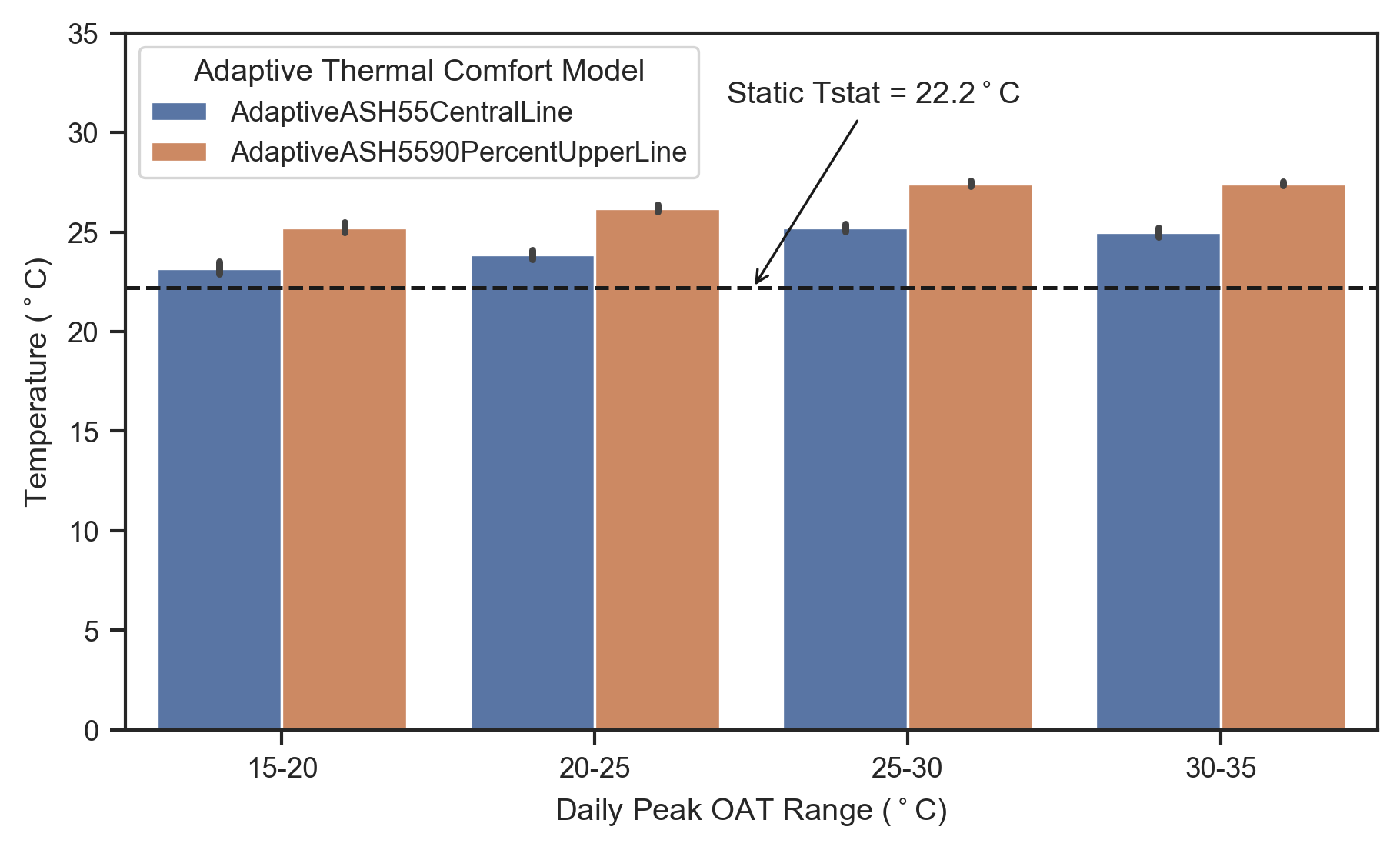}}
\caption{Comparison of Thermostat Cooling Setpoints between the Static and Adaptive Comfort Models}
\label{fig2}
\end{figure}

The cooling setpoints in climate zone 4A, based on both Adaptive-55 and Adaptive-90, are shown in Figure \ref{fig2}. The X-axis is the maximum daily outdoor air temperature, and the Y-axis is the mean cooling setpoint temperature. The baseline static cooling setpoint at 22.2°C is plotted as a dashed line. In climate 4A, the adaptive cooling setpoint based on Adaptive-55 is never above 25.8°C and the adaptive cooling setpoint based on Adaptive-90 is never above 28.3°C.

\begin{figure}[htbp]
\centerline{\includegraphics[width=1.0\columnwidth]{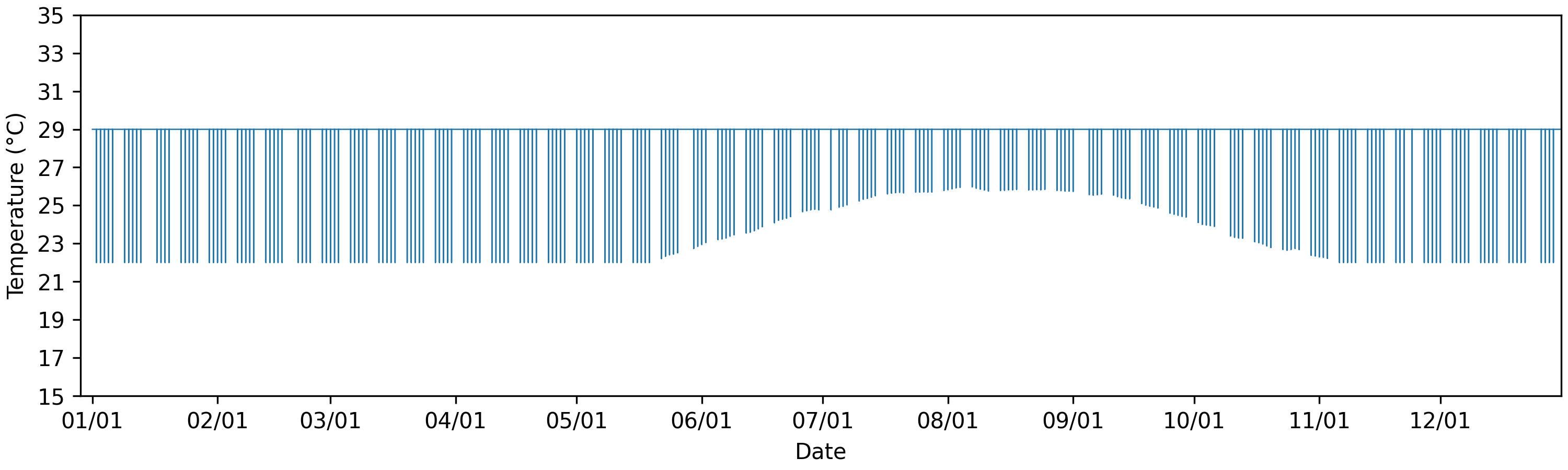}}
\caption{Annual Cooling Setpoints}
\label{fig3}
\end{figure}

The daily cooling setpoints in climate zone 4A based on Adaptive-55 are shown in Figure \ref{fig3}. The cooling setpoint changes in the same direction of prevailing outdoor temperature, which starts to rise from late May, peaks in August, and falls back to the baseline setback temperature in November. On weekends and holidays, the system cooling setpoint is the same as the setback temperature at 29°C. 

\begin{figure}[htbp]
\centerline{\includegraphics[width=1.0\columnwidth]{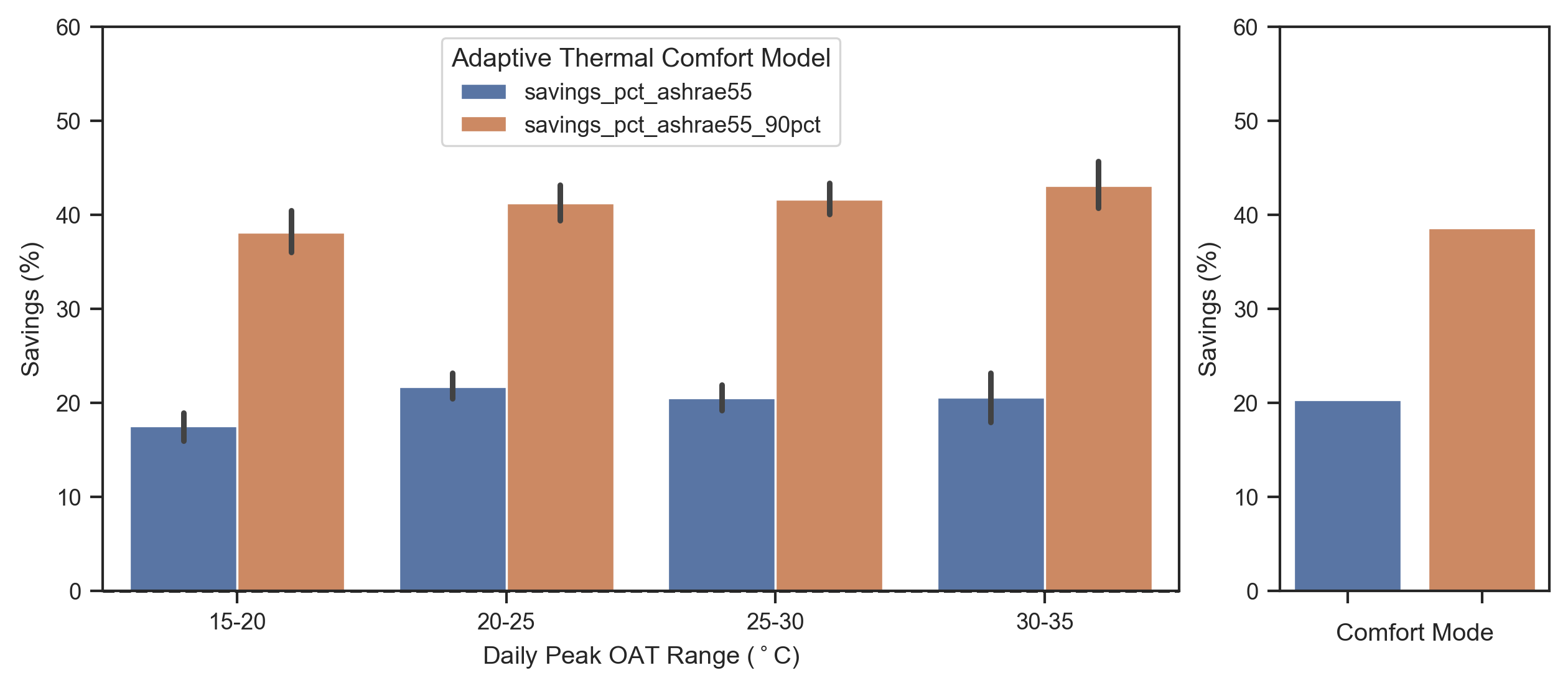}}
\caption{Comparison of HVAC Savings (\%) between ASHRAE55-CentralLine and 90PercentUpperLine
under different daily weather conditions (left) and total savings in climate 4A (right) }
\label{fig4}
\end{figure}

Raising cooling setpoints by a mere few degrees from the baseline leads to significant energy savings. As shown in Figure \ref{fig4}, the cooling setpoints based on Adaptive-55, ranging from 22.2°C to 25.8°C, achieve approximately 20\% of total cooling energy savings; the cooling setpoints based on Adaptive-90, ranging from 22.2°C to 28.3°C, achieve an even higher 38\% of total cooling energy savings. This cooling energy savings include both compressor energy savings and fan energy savings. 

\subsection{Occupancy-based HVAC and Plug Load Control}
\begin{figure}[htbp]
\centerline{\includegraphics[width=1.0\columnwidth]{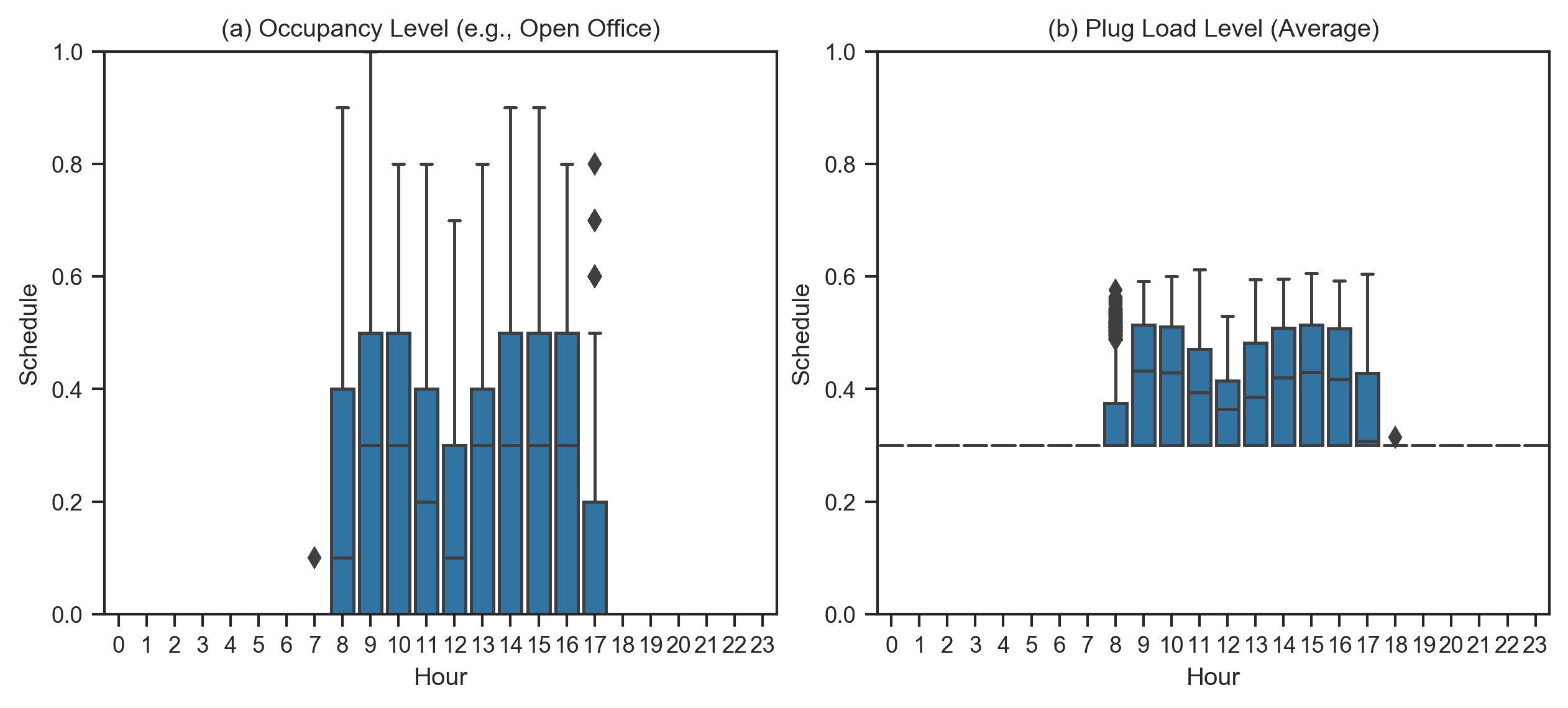}}
\caption{Dynamic Occupant Behaviours (e.g., Open Office) and Plug Load on Weekdays}
\label{fig5}
\end{figure}

The occupancy input is based on the realistic schedule for office space developed at Lawrence Berkeley National Laboratory (LBNL) by incorporating real data collection and stochastic modeling methods \cite{Chen2018AnSimulator}. It is also the result of the International Energy Agency Energy in Buildings and Communities Programme (IEA-EBC) Annex 66 - Definition and Simulation of Occupant Behavior in Buildings \cite{IEA-EBCBuildings}. The realistic occupancy level at 10-minute intervals is generated by LBNL’s occupant behavior web application \cite{OccupancySimulator} considering the diversity and stochastics of occupant activities. The occupancy level is shown in Figure \ref{fig5}. Unlike the traditional assumption of static occupancy e.g. 8 am - 5 pm, the stochastic occupancy takes into account the variations in real offices setting, providing a high-resolution occupancy level at sub-hourly intervals.  As shown in Figure \ref{fig5}, the occupancy level in the early morning, around noon, and in the evening has larger variations due to day-to-day and person-to-person differences. 

The accurate occupancy sensing enabled by IoT sensors not only saves HVAC system energy consumption but also helps to save electricity on lighting and plug load. The smart plug control automatically turns off non-critical equipment and appliances (such as monitors, televisions, coffee makers, water heaters, etc.) if occupancy is not detected for an extended period. The occupancy-centric control alone achieves 7.7\% of total electricity consumption savings in small offices located in Climate 4A, by reducing 6.7\% HVAC consumption and 28.2\% plug loads. The electricity intensity is thus reduced from 132.4 $kWh/m^2$ to 122.2 $kWh/m^2$, achieving substantial savings of 10.2 $kWh/m^2$.

\begin{figure}[htbp]
\centerline{\includegraphics[width=1.0\columnwidth]{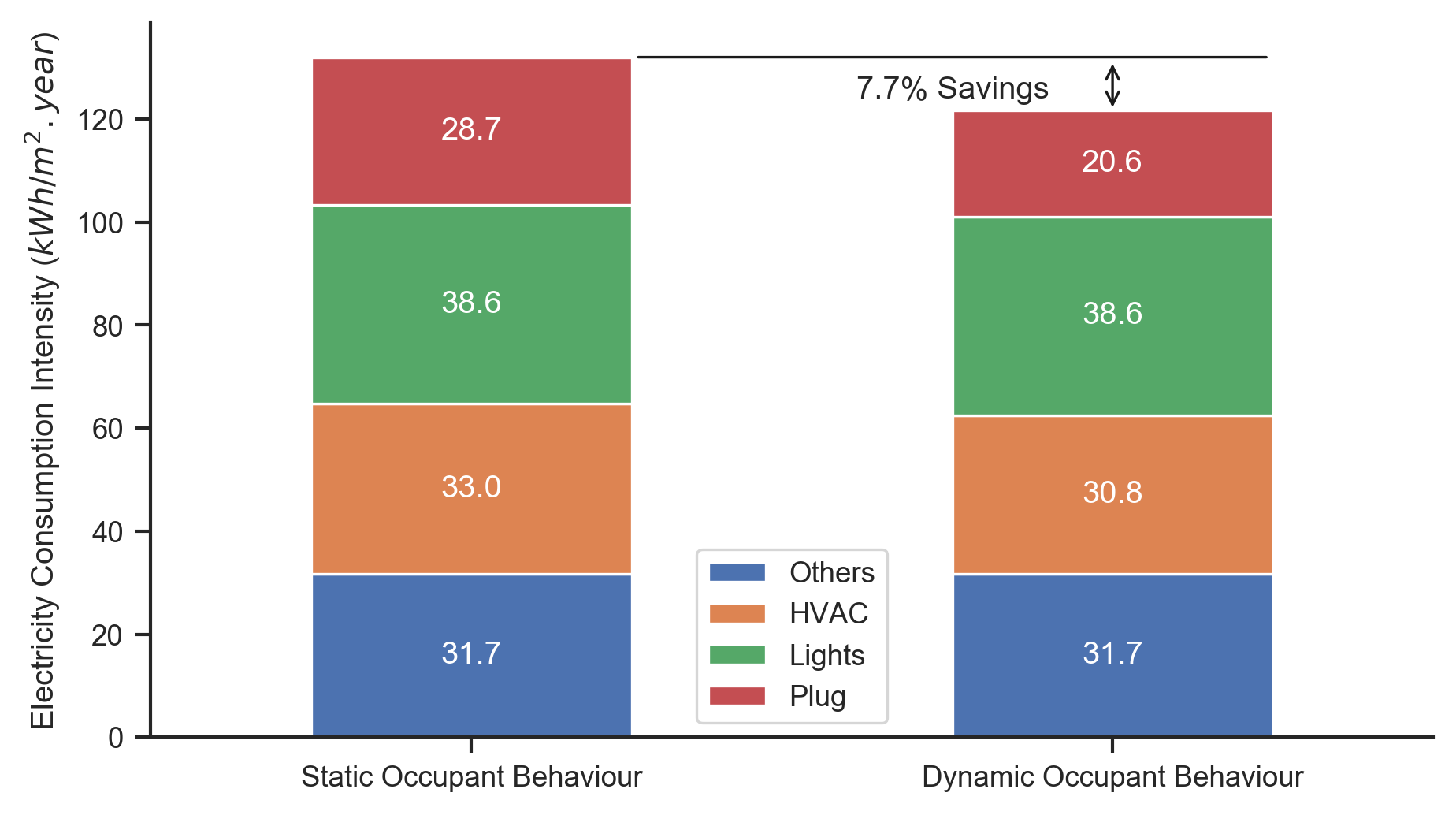}}
\caption{Comparison of Electricity Consumption Intensity between Static and Dynamic Occupancy-based Control for Small Offices in Climate 4A}
\label{fig6}
\end{figure}

\subsection{Passive Cooling Strategies (Night Purge)}
We implement a strategy for precooling a building at night using outdoor air. This strategy can consist of running the system fans with the outdoor air dampers open when outdoor conditions are favorable for a precooling strategy. The zone terminal unit air dampers may also be held at their fully open position to minimize fan energy consumption while precooling.

\begin{figure}[htbp]
\centerline{\includegraphics[width=1.0\columnwidth]{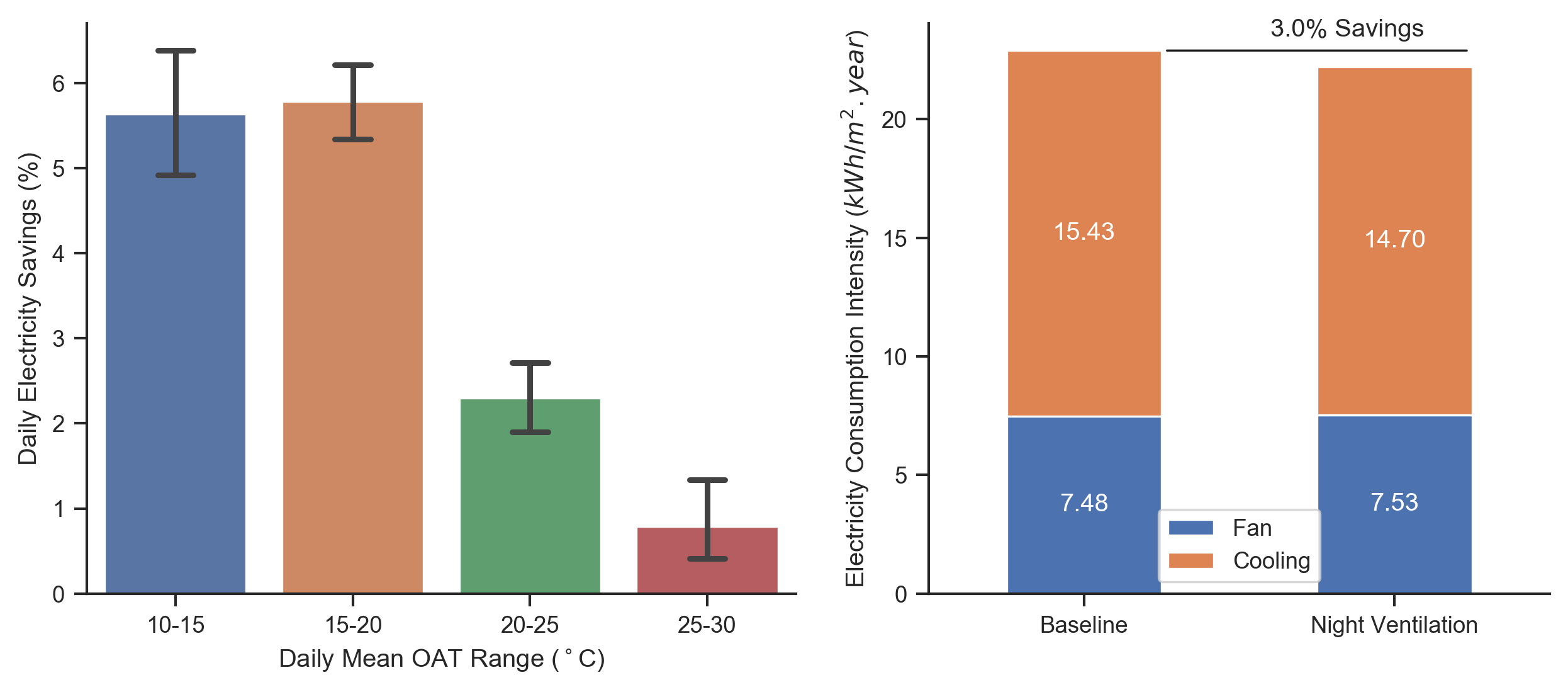}}
\caption{Daily Energy Saving from HVAC by Different Daily Mean OAT Ranges (left) and Cooling Electricity Consumption intensity in Cooling Season (right)}
\label{fig7}
\end{figure}

Figure \ref{fig7} shows the daily energy saving from the HVAC by different daily mean OAT ranges for a small office building in the climate zone 4A. It can be seen that the night purge strategy has played a role in reducing HVAC energy consumption when the climate is suitable. When the daily mean OAT exceeds about 25 ºC, night ventilation may not be able to effectively reduce the HVAC operating time during the day, thus wasting the energy of the fan at night.

During the cooling season, the night purge alone saves 3.0\% of HVAC energy consumption in small offices located in Climate 4A, as shown in Figure \ref{fig7}. The energy intensity is reduced by approximately 0.7 $kWh/m^2$, which includes 0.7 $kWh/m^2$ of cooling energy savings and 0.05 $kWh/m^2$ of fan energy increase.

% \begin{figure}[htbp]
% \centerline{\includegraphics[width=1.0\columnwidth]{figures/fig8-night-ventilation-comparison-4A.png}}
% \caption{Effect of Night Ventilation on Reducing Energy Consumption of HVAC in Cooling Season}
% \label{fig8}
% \end{figure}

Night purge is most effective in the buildings with high thermal mass and climates that diurnal temperature swing is significant or summer is mild \cite{Fernandez2017ImpactsReduction}. In the cases when buildings have exposed concrete slabs, the night purge will cool down the slab temperature, which will help reduce the daytime HVAC energy consumption thanks to lowered radiant temperature and operative temperature. When night purge is applied along with phase changing material and adaptive thermal comfort setpoint, it will create a synergistic effect and unlock more energy savings \cite{Solgi2016CoolingVentilation}. The cooling demand on the HVAC system is greatly reduced as the phase-changing material that is cooled by the cool night air, slowly absorbs heat throughout the day and keeps indoor air temperature close to the adaptive thermal comfort temperature.

\subsection{Combined Measures}
\begin{figure}[htbp]
\centerline{\includegraphics[width=1.0\columnwidth]{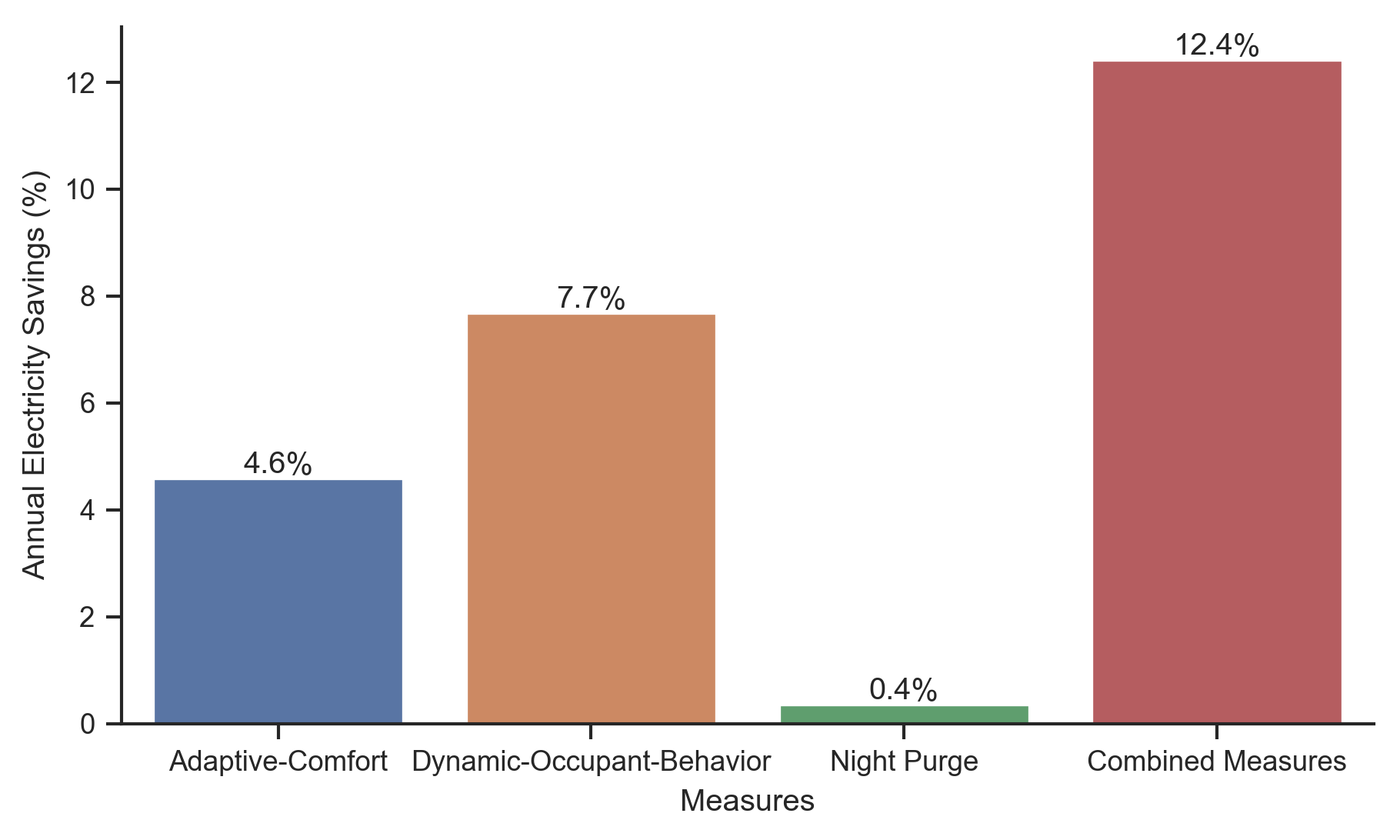}}
\caption{Potential Annual Electricity Savings for Individual and Combined Measures in Small Office}
\label{fig9}
\end{figure}

Combining the three optimization strategies above, buildings can achieve the highest energy savings. As shown in Figure \ref{fig9}, the standard small office in climate zone 4A reduces annual electricity consumption by 12.4\%, implementing adaptive setpoint, occupancy-centric control, and night purge natural ventilation. Among these individual measures, occupancy-centric control is the most effective strategy and leads to a 7.7\% reduction in total electricity consumption, as the savings come from the HVAC system, plug load, and lighting. The adaptive setpoint achieves a 4.6\% reduction in annual electricity savings, all of which stems from reduced cooling energy of the HVAC system. The night purge ventilation is the least effective measure and saves 0.4\% of total electricity consumption, which is also solely from reduced cooling energy consumption. The interactive effect of these three measures is insignificant as the sum of individual savings (12.7\%) is very close to the combined savings (12.4\%). 

\subsection{Comparison between Different Climates}
\begin{figure}[htbp]
\centerline{\includegraphics[width=1.0\columnwidth]{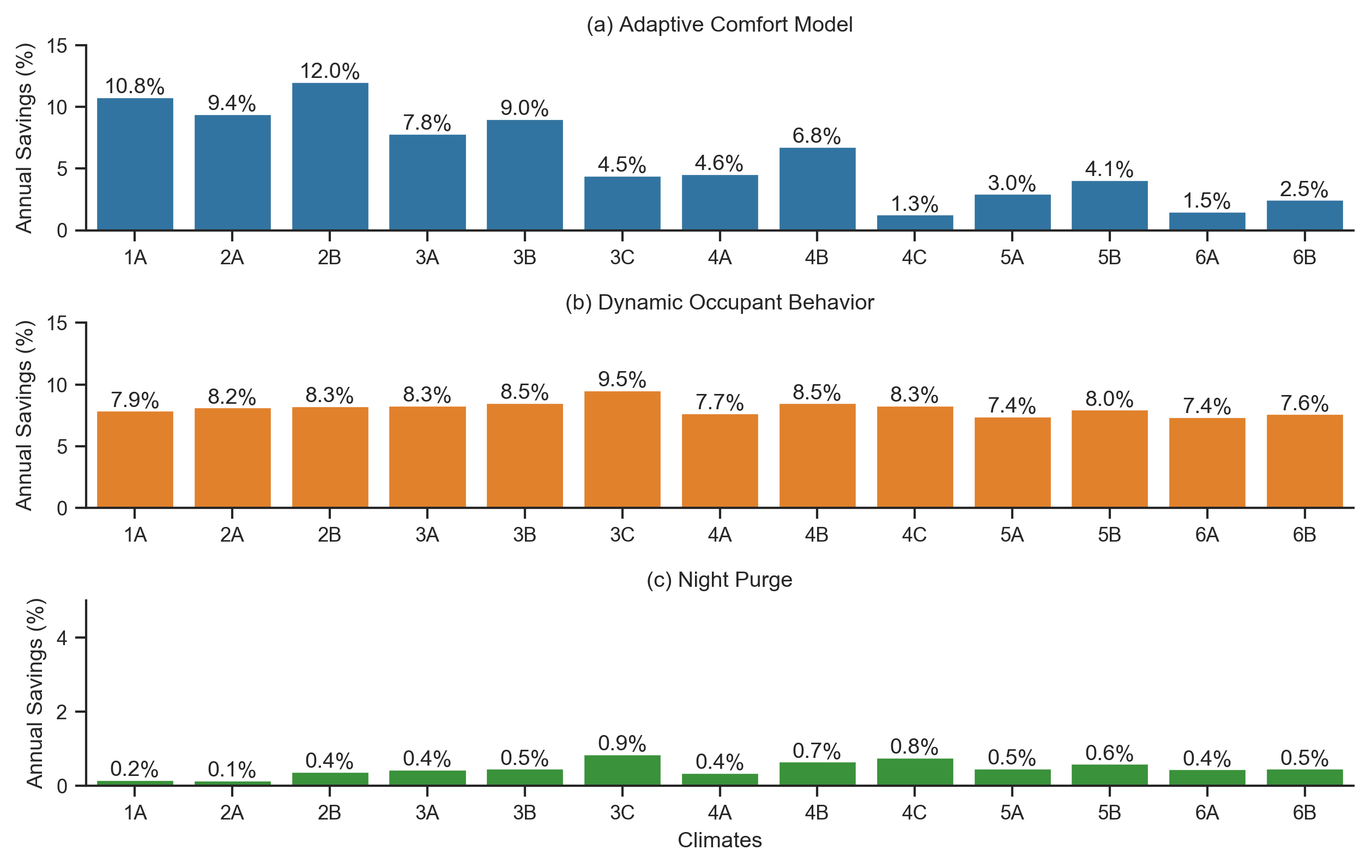}}
\caption{Comparison of Energy Efficiency Measures between Different Climates}
\label{fig10}
\end{figure}

Figure \ref{fig10} shows the annual electricity savings comparison of EE measures between different climates. It can be clearly seen that the effect of the dynamic occupancy behavior model is consistent in all climate conditions, saving about 7.4-9.5\% per year. In contrast, the adaptive comfort model is primarily influenced by the climate, as shown in Figure \ref{fig10} (a). In terms of reducing cooling power consumption, the effect of the adaptive comfort model decreases from 12\% to 1.5\% as the climate changes from hot to cold. On the other hand, the humidity also has a large impact on the effect of the adaptive comfort model. Under the same climatic conditions, a drier climate can achieve much higher savings compared to a humid climate.

Among the three measures, the effect of the "night purge" strategy is fairly limited, saving only about 0.1\% to 1\% per year. As shown in Figure \ref{fig10} (c), the most effective climate zone for night purge is 3C (ocean), which is warm during the day and cool at night. Previous studies have shown that in dry climates with warm days and cool nights, this measure is usually worth considering.

\begin{figure}[htbp]
\centerline{\includegraphics[width=1.0\columnwidth]{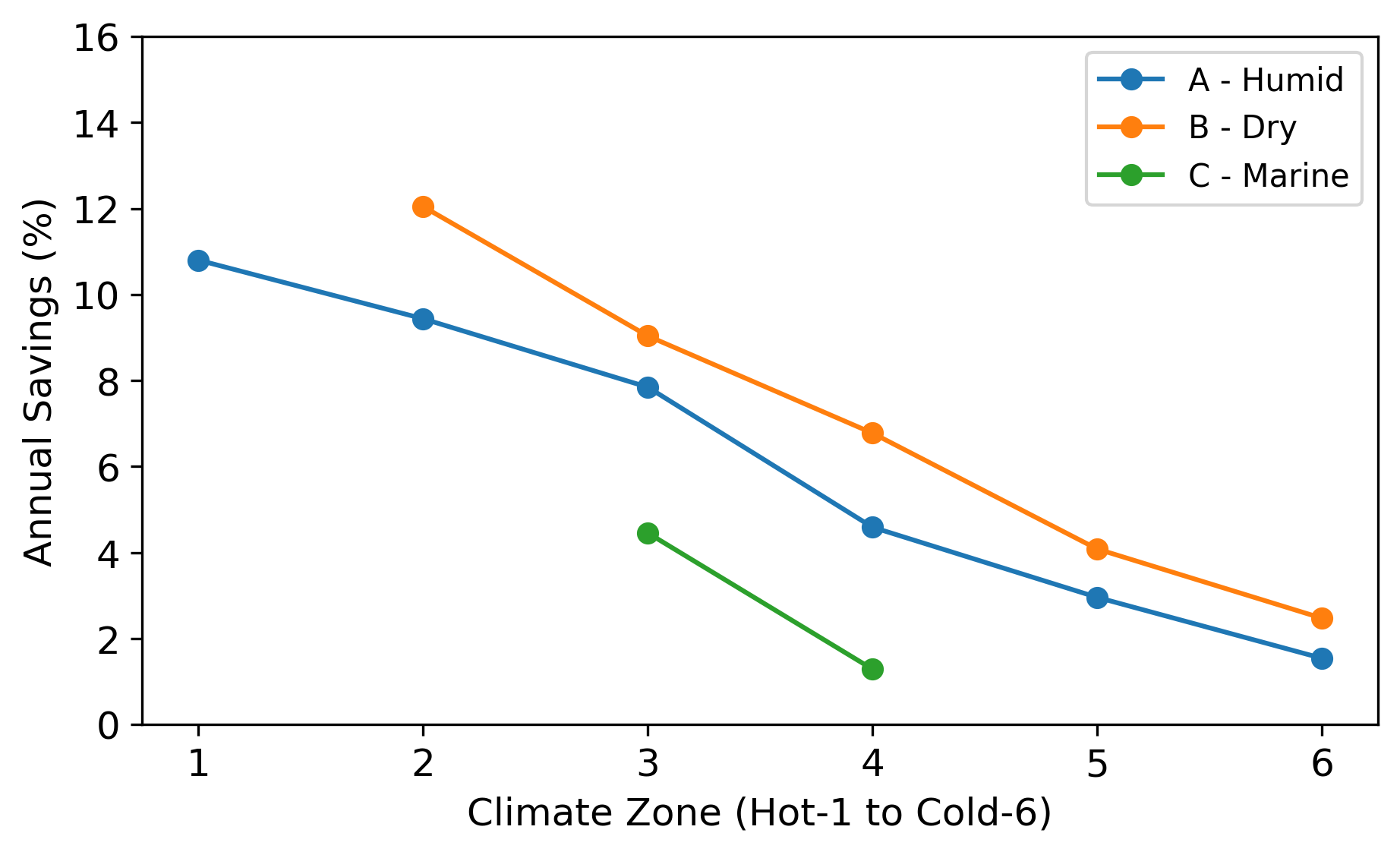}}
\caption{Cooling Energy Savings by Climates}
\label{fig11}
\end{figure}

From hot to cold climates, the annual electricity savings percentage by adopting adaptive thermal comfort decreases as the cooling energy is responsible for smaller proportions of total electricity consumption in cooler climates. Climates that belong to different humidity subcategories also show different levels of cooling energy savings. As shown in Figure \ref{fig11}, dry climate, or subcategory B, benefits most from the adaptive cooling setpoints. Humid climate, or subcategory A,  sees a lower saving percentage as the HVAC system consumes energy to meet the latent cooling load for moisture removal. Marine climate, or subcategory C, shows the least saving percentage because of its mild summer that results in a smaller percentage of cooling energy of the total electricity consumption. 

\begin{figure}[htbp]
\centerline{\includegraphics[width=1.0\columnwidth]{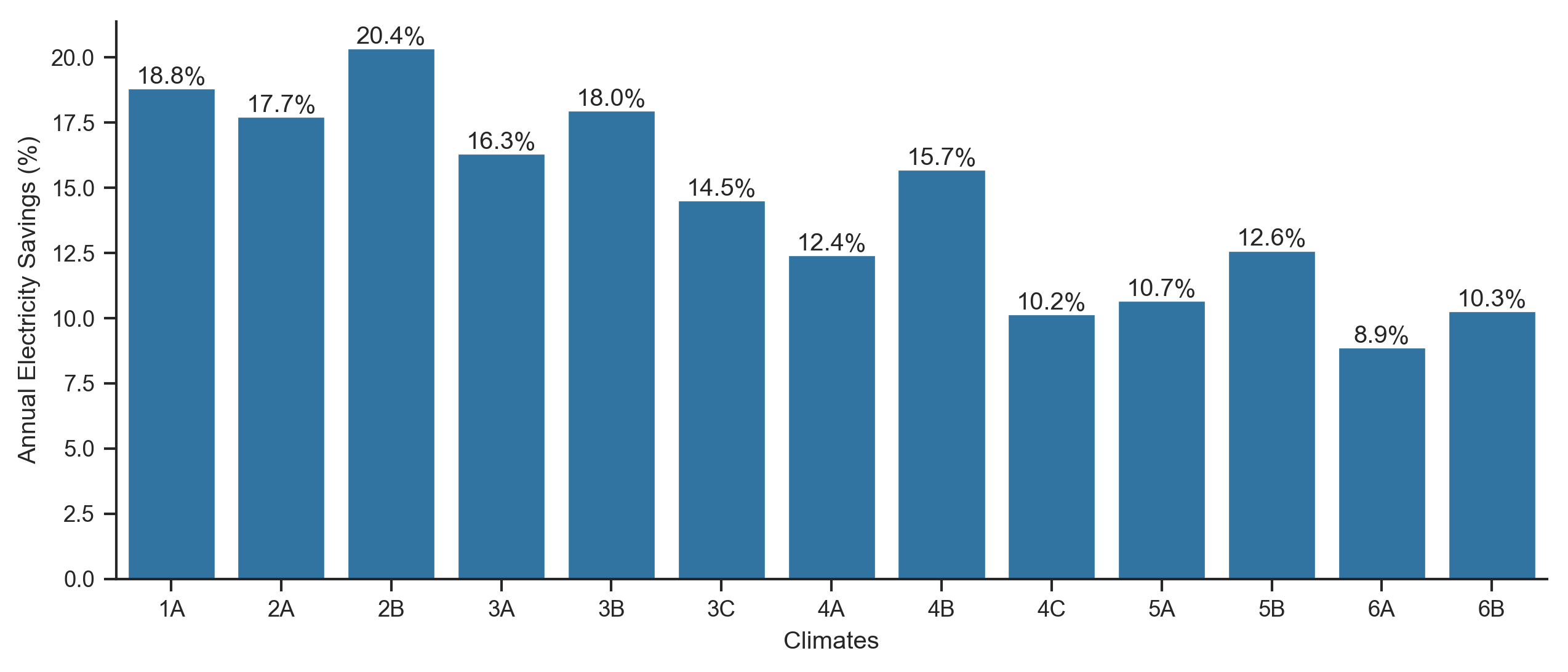}}
\caption{Potential Annual Electricity Savings for Combined Measures in Small Office among Difference Climates}
\label{fig12}
\end{figure}

Figure \ref{fig12} presents the annual electricity savings achieved by combining three EE measures across all climate zones in the United States. In the hot and dry climate zone "2B", the highest saving is about 20\%. In contrast, the lowest saving observed in the cold climate zone “6A” is only 8.9\%, and most of the electricity savings come from smart plug load control.

Table \ref{tab3} summarizes the annual electricity savings at the entire building level and HVAC end-use level in different climate regions in the United States. The deployment of combined energy efficiency measures can save 12.2 $kWh/m^2$ to 30.4 $kWh/m^2$ per year. Most of the savings come from HVAC end-use, which is about 6.5 $kWh/m^2$ to 22.6 $kWh/m^2$. Compared to the baseline, combined energy efficiency measures can achieve electricity savings of 9\% to 20\% for the entire building, and 23\% to 39\% for HVAC end-use. 

\begin{table}[htbp]
\caption{Annual Electricity Saving of Combined Energy Efficiency Measures for Small Offices in Different Climates}
\begin{center}
\begin{tabular}{|c|c|c|c|c|}
\hline
\multirow{2}{*}{\textbf{Climate Zone}} & \multicolumn{2}{c|}{\textbf{Annual Savings ($kWh/m^2$)}} & \multicolumn{2}{c|}{\textbf{Annual Savings (\%)}} \\ \cline{2-5} 
                                        & \textbf{Total Electricity}           & \textbf{HVAC}           & \textbf{Total Electricity}         & \textbf{HVAC}         \\ \hline
1A                                      & 29.5                        & 21.4                    & 19\%                      & 33\%                  \\ \hline
2A                                      & 26.6                        & 18.6                    & 18\%                      & 32\%                  \\ \hline
2B                                      & 30.4                        & 22.6                    & 20\%                      & 39\%                  \\ \hline
3A                                      & 22.3                        & 15.0                    & 16\%                      & 36\%                  \\ \hline
3B                                      & 25.1                        & 17.6                    & 18\%                      & 37\%                  \\ \hline
3C                                      & 18.3                        & 10.5                    & 15\%                      & 30\%                  \\ \hline
4A                                      & 16.4                        & 10.0                    & 12\%                      & 30\%                  \\ \hline
4B                                      & 21.1                        & 13.9                    & 16\%                      & 35\%                  \\ \hline
4C                                      & 12.4                        & 6.6                     & 10\%                      & 25\%                  \\ \hline
5A                                      & 14.3                        & 7.7                     & 11\%                      & 29\%                  \\ \hline
5B                                      & 16.9                        & 10.2                    & 13\%                      & 30\%                  \\ \hline
6A                                      & 12.2                        & 6.5                     & 9\%                       & 23\%                  \\ \hline
6B                                      & 13.4                        & 7.2                     & 10\%                      & 26\%                  \\ \hline
\end{tabular}
\label{tab3}
\end{center}
\end{table}
\include{sec4-discussion}
\section{Conclusion}

The U.S. commercial buildings have substantial energy-saving potential if the building control could address varying occupancy, occupant’s thermal preference, and utilizing free cooling. This study quantifies the energy savings in small office buildings across all major U.S. climate zones, based on the commercial prototype building models developed by the U.S. Department of energy. The energy-saving effectiveness of each individual measure and all measures combined are presented.

The adaptive cooling setpoints are applied to avoid over-cooling in commercial buildings, taking into consideration people's adaptive behavior and clothing adjustment against general weather conditions. This measure is extremely effective in reducing cooling energy consumption, especially in climate 2B, in which the adaptive cooling setpoint alone saves 12\% electricity consumption of the small office building, compared to the baseline static cooling setpoint at 22.2°C. 

The occupancy-centric control, enabled by IoT sensors that detect occupancy and motion, is applied to both the HVAC system and smart plugs control. The corresponding savings of the HVAC system energy consumption is around 6.7\% and savings of plug load is about 28.2\%. This measure alone saves approximately 7.7\% of total electricity consumption across all climate zones. 

The night purge ventilation utilizes free cooling and improves indoor air quality. This measure alone reduces up to 3.0\% of the cooling energy consumption of the HVAC system for small offices in climate 4A. Specifically, it is most effective in marine climates such as 3C (San Francisco) and 4C (Seattle), and can achieve approximately 5.6\% of cooling electricity consumption. It is equivalent to about 0.9\% of the total electricity consumption of the building. 

Significant total electricity savings ranging from 8.9\% to 20.4\% can be achieved by combining all three measures, which are equivalent to 12.2 $kWh/m^2$ to 30.4 $kWh/m^2$ annual reductions in electricity usage intensity. The electricity-saving percentages show a decreasing trend from hot to cold climate. For climate zones with similar numbers of cooling degree days and heating degree days, the dry subzones achieve the highest energy savings.

Enabled by building IoT sensors, smart hardware, and automated building subsystems, more energy-saving opportunities can be unlocked by smart control to improve the energy efficiency of commercial buildings. This study quantifies the electricity savings from adaptive setpoint, occupant-centric control, and night purge ventilation in small office buildings across all U.S. climate zones. The next step is to carry out experiments of such "plug and play" IoT sensing and control devices (for example, smart thermostats with occupancy sensors, smart plugs, window actuators) in small offices and quantify the energy-saving and demand reduction potential.



\bibliographystyle{elsarticle-num}
\bibliography{references.bib}






% \end{thebibliography}

\end{document}